# The AI Interface: Designing for the Ideal Machine-Human Experience
Editorial[1]

Aparna Sundar, Tony Russell-Rose, Udo Kruschwitz, Karen Machleit

## Abstract

As artificial intelligence (AI) becomes increasingly embedded in daily life, designing intuitive, trustworthy, and emotionally resonant AI-human interfaces has emerged as a critical challenge. This editorial introduces a Special Issue that explores the psychology of AI experience design, focusing on how interfaces can foster seamless collaboration between humans and machines. Drawing on insights from diverse fields—healthcare, consumer technology, workplace dynamics, and cultural sectors—the papers in this collection highlight the complexities of trust, transparency, and emotional sensitivity in human-AI interaction. Key themes include designing AI systems that align with user perceptions and expectations, overcoming resistance through transparency and trust, and framing AI capabilities to reduce user anxiety. By synthesizing findings from eight diverse studies, this editorial underscores the need for AI interfaces to balance efficiency with empathy, addressing both functional and emotional dimensions of user experience. Ultimately, it calls for actionable frameworks to bridge research and practice, ensuring that AI systems enhance human lives through thoughtful, human-centered design.

## Background and motivation

As artificial intelligence continues to integrate more deeply into our daily lives, the design of AI interfaces has become pivotal in shaping the quality of human-machine interactions. The challenge is not just technological but deeply emotional, ethical, and cognitive. Across various fields—from healthcare and consumer products to workplace dynamics and cultural sectors—the focus has shifted to developing AI technologies to deploying systems that promote trust, empathy, and seamless cooperation between humans and machines.

The aim of this Special Issue is to deepen our understanding of the psychology of design, particularly in how we approach and create AI interfaces such as chatbots, robotics, IoT devices, AI assistants, and more. Artificial intelligence (AI) and machine learning (ML) are transformative technologies that organizations worldwide are increasingly investing in. AI can replicate tasks requiring human intelligence by leveraging probabilistic outcomes based on real-world data to predict future scenarios. ML processes vast amounts of data to develop and validate decision-making logic, often inspired by biological neuron signals, as seen in deep learning or natural language processing (NLP). Additionally, no-code tools now enable business analysts to make ML predictions without prior expertise. This Special Issue examines the role of AI in shaping human perception and inference, with a focus on designing the ideal machine-human experience.

---
[1] https://www.sciencedirect.com/special-issue/10XWFHP2WVB

User perception can vary widely due to individual differences, environmental factors, and cultural influences, all of which impact user experiences. User inferences are the mental processes that enable individuals to draw conclusions, make judgments, and generate new knowledge based on the information they encounter. Together, perception and inferences shape the overall user experience. In AI, designing systems that are transparent, intuitive, and aligned with users' needs and expectations is critical. With the rapid growth of technological investment and innovation in this field, the need for research has become even more pressing—particularly from a design and product development perspective. User experience research plays a key role in bridging the gap between AI and its users. Beyond basic usability, it addresses critical factors such as user mental models, trust, and transparency. This includes exploring how psychology can inform the design of machine-human interfaces, as well as the personalization and adaptability of AI assistants, among other considerations.

One of the most critical aspects of interacting with AI is the language it uses in messaging or persuasion attempts. Despite its importance, we know relatively little about the psychology of AI experience design. Research on how marketers communicate with consumers highlights robust effects of factors like message tone (Sundar & Cao, 2018; Sundar & Paik, 2017), message repetition (Sundar, Kardes & Wright, 2015), and language structure and categorization (Schmitt & Zhang, 1998). While scholars have explored the anthropomorphic relationships individuals form with AI assistants (Uysal, Alavi, & Benzencon, 2022), how people react to and respond to AI remains underexplored. This Special Issue seeks to bridge disciplines such as communication, marketing, and judgment and decision-making to better understand these dynamics. Research in this area aims to uncover how humans perceive communication, particularly when claims originate from AI assistants or sources not perceived as human.

This Special Issue explores the evolving relationship between AI, ML, and human behavior. It features articles that delve into the complexities of AI-human interactions, potential human perceptions, and insights to enhance machine learning and inferences on both sides of the interaction—paving the way for meaningful technological advancements. Highlighting the role of AI and ML in the digital evolution of computing, this issue emphasizes the psychological and human responses to transformative digital technologies.

What does the ideal AI-human interface look like? Conventional UX wisdom would suggest the answer is, "It depends." And indeed, it does—on the use case. The variety of papers in this collection illustrates that the ideal interface varies as widely as the applications themselves. Take ChatGPT, or its predecessor GPT-3, released in 2020. Dale (2024) notes that "the simple genius of wrapping up the technology in a chat interface made it easily available to every internet user, regardless of their technical expertise beyond conventional browser use." Clearly, for that purpose, the ideal interface is one that prioritizes accessibility and simplicity—a design that would not translate to interactions with autonomous vehicles. As new challenges arise in human-AI interaction, they demand innovative interfaces and approaches to address them (e.g., Ognibene et al., 2022).

Simplicity is not always the best measure for assessing human-computer interaction (Sarkar, 2023). This collection reinforces that finding, highlighting, for example, the vital role of trust (Mayer et al., 2024). Trust is a cornerstone of human-AI interaction, so it is no surprise that it emerges as a key theme across this collection of research papers. This is not a new concept—adoption of AI-enabled systems has long been closely tied to trust (Bach et al., 2022), even before the rise of chat interfaces for Large Language Models (Denning, 2023; Church, 2024). Leschanowsky et al.'s systematic literature review on trust perception, security, and privacy in conversational AI contributes significantly to understanding these issues from a user perspective.

A somewhat puzzling finding is reported by Hu et al. (2024), who reveal that full transparency of algorithmic decisions in AI interactions is not always a key requirement. This highlights the multi-faceted nature of human-AI interaction, contrasting starkly with use cases where a lack of transparency significantly undermines trust—such as the controversy surrounding the opaque COMPAS system used to decide parole outcomes[2]. The paradox reported in Hu et al.'s paper exemplifies the inherent complexity of the research problems that inspired this Special Issue. It is not an isolated case, as similar findings are reflected in user studies by Aslett et al. (2024). Their research, exploring the impact of deploying a search engine to evaluate the trustworthiness of online information, concluded that "the strategy of pushing people to verify low-quality information online might paradoxically be even more effective at misinforming them."

This collection of studies and surveys advances our understanding of how to approach the design of human-AI interaction. While it represents a snapshot of the field, it also serves as a foundation for future research. The results reported in the eight accepted papers offer valuable insights to drive the state of the art forward. However, the field is evolving rapidly. Progress in AI continues to reshape our understanding of human-AI interaction, even in mature domains such as search engines. Well-established design patterns (Russell-Rose and Tate, 2012) are being redefined, as evidenced by the shift towards AI agents (White, 2024). Perhaps the most transformative development is the emergence of Generative AI (GenAI), described as a paradigm shift with profound implications. Traditional methods of developing, evaluating, and deploying AI systems are being fundamentally challenged (Miikkulainen, 2024). The inclusion of several review papers in this collection allows readers to reflect on the rapidly changing research landscape—a landscape producing new findings at an unprecedented pace.

---

[2] https://www.nytimes.com/2017/05/01/us/politics/sent-to-prison-by-a-software-programs-secret-algorithms.html

## Insights from this Issue

Affective communication forms the foundation of an ideal AI-human experience, as explored in Affective Foundations in AI-Human Interactions (Liu et al., 2024). The researchers argue that for AI to evolve beyond mere task execution, it must connect with human emotions in meaningful ways. Drawing from evolutionary biology and interspecies communication, they propose models for AI that resonate with universal affective pathways. These insights underline the necessity of moving beyond human-like cognition toward a design where AI can genuinely empathize with its users, establishing trust and intuitive interaction. This idea of an emotional bridge between humans and machines is critical for envisioning future AI interfaces.

Yet, trust is a double-edged sword, as highlighted in Evaluating Privacy, Security, and Trust Perceptions in Conversational AI (Leschanowsky et al., 2024). Even as conversational AI systems, such as voice assistants, become more integrated into daily life, users remain wary about privacy, security, and trust. The study emphasizes the importance of reliable, validated metrics for gauging these concerns, urging designers to consider not only how AI interfaces perform but also how they are perceived in terms of safety and privacy. This connection between emotional resonance and trust speaks to a broader need for AI interfaces that respect user boundaries while offering personalized interactions.

As AI systems become increasingly embedded in our daily lives, transparency in their operations becomes crucial. However, The Transparency-Resistance Paradox in Algorithmic Management (Hu et al., 2024) reveals the complexities of this challenge in algorithm-driven gig work. While transparency may initially improve worker satisfaction, too much transparency can lead to resistance, complicating the AI-human relationship. This paradox raises an important point: designing AI interfaces for the workplace must balance efficiency with human factors like fairness and empathy, suggesting that the ideal machine-human interface may not always be hyper-transparent, but rather sensitive to human emotional thresholds. The proposed model of "Human-Algorithm Co-management" reflects this, where AI's cold logic is tempered by human empathy, reinforcing that an optimal interface must integrate human warmth with machine precision.

Consumer technology interfaces also face similar challenges, particularly in framing and presenting AI to users. When Being Smart Trumps (Smale et al., 2024) explores consumer preferences for products labeled "smart" versus "AI-enabled," finding that people are more receptive to "smart" products due to the anxiety that AI labels provoke. This highlights a critical insight for AI interface design: how AI is framed and communicated can significantly affect user acceptance. For AI systems to succeed in consumer markets, designers must reduce anxiety through careful labeling and interaction design, framing AI capabilities in ways that enhance user comfort.

Healthcare offers another vital context for examining machine-human interfaces, where trust and emotional sensitivity are paramount. In Evaluating the Interactions of Medical Doctors with Chatbots, Triantafyllopoulos et al. (2024) report mixed reactions from doctors using ChatGPT for medical consultations. While younger doctors appreciated the system's responsiveness, more experienced physicians expressed dissatisfaction, particularly concerning clarity and accuracy. This finding underscores the need for AI interfaces in healthcare to be finely tuned to professional expertise levels and context-specific requirements. Similarly, Attitudes towards Digitalized and AI-Based Medical Consultations (Mayer et al., 2024) finds that while AI-based consultations can improve efficiency, trust and user preference decline when the conversation involves sensitive or emotional topics, such as receiving bad news. These studies collectively emphasize that, in healthcare, the ideal AI interface must not only provide accurate information but also recognize the emotional weight of interactions, particularly in sensitive medical contexts.

The broader challenge of integrating AI into emotionally charged domains extends to overcoming bias. To Err is Human: Bias Salience Can Help Overcome Resistance to Medical AI (Isaac et al., 2024) presents an innovative strategy to address skepticism toward medical AI by highlighting human biases. When users recognize the limitations of human decision-making, they are more inclined to trust AI as a neutral alternative. This insight holds significance for AI interface design, as it suggests that making users aware of AI's strengths—its fairness and impartiality—can help reduce resistance and foster trust in machine-driven decision-making, particularly in areas like healthcare where trust is crucial.

Lastly, the cultural sector provides a unique lens to understand how AI interfaces can enhance user experiences in traditionally human-centric environments. A Systematic Review of Digital Transformation Technologies in Museum Exhibition by Jingjing Li et al. (2024) shows how AI and other digital transformation technologies are revolutionizing museum exhibitions. These technologies offer new ways to engage audiences, enhance accessibility, and collect visitor feedback. However, as the study notes, the connection between these technologies and user experience remains underexplored. This points to an opportunity for AI designers to bridge the gap between advanced technologies and human-centered experiences, ensuring that AI enhances rather than detracts from cultural appreciation.

## Conclusions

Across these diverse fields, the narrative remains clear: the design of AI interfaces must go beyond efficiency and functionality. Whether in consumer products, healthcare, workplace management, or cultural heritage, the ideal AI interface fosters emotional connection, balances transparency with trust, and carefully frames the AI's role to reduce anxiety. The future of AI-human interactions depends on designing interfaces that not only perform but also resonate emotionally, acknowledging human needs for empathy, fairness, and intuitive communication. The challenge lies in ensuring that these AI systems, while powerful and advanced, remain fundamentally human in their touch.

While this collective body of research presents valuable insights and directions, there is still work to be done in providing the kind of practical, hands-on guidance that UX designers and product managers need to implement these ideas effectively. Theoretical models and high-level findings offer a necessary foundation, but the next phase in AI interface design must focus on translating these insights into concrete, actionable strategies. Industry professionals are seeking tools, frameworks, and best practices that can be applied directly to product development—making this a critical gap that needs addressing. As AI becomes more integral to everyday experiences, bridging this gap between research and practical application will be essential in creating AI systems that truly enhance human lives.

## References:

Aslett, K., Sanderson, Z., Godel, W. *et al.* (2024). Online searches to evaluate misinformation can increase its perceived veracity. *Nature* 625, 548–556.
https://doi.org/10.1038/s41586-023-06883-y

Bach, T. A., Khan, A., Hallock, H., Beltrão, G., & Sousa, S. (2022). A Systematic Literature Review of User Trust in AI-Enabled Systems: An HCI Perspective. *International Journal of Human–Computer Interaction*, *40*(5), 1251–1266.
https://doi.org/10.1080/10447318.2022.2138826

Cabrera, Á. A., Tulio Ribeiro, M., Lee, B., Deline, R., Perer, A., & Drucker, S. M. (2023). What did my AI learn? how data scientists make sense of model behavior. *ACM Transactions on Computer-Human Interaction*, *30*(1), 1-27, https://dl.acm.org/doi/pdf/10.1145/3542921.

Church, K. (2024). Emerging trends: When can users trust GPT, and when should they intervene? *Natural Language Engineering*. 30(2):417-427. doi:10.1017/S1351324923000578

Dale, R. (2024). A year's a long time in generative AI. *Natural Language Engineering*, 30(1):201-213. doi:10.1017/S1351324923000554

Denning, P. J. (2023). Can Generative AI Bots Be Trusted? Communications of the ACM 66, 6 (June 2023), 24–27. https://doi.org/10.1145/3592981

Gao, Y., & Liu, H. (2022). Artificial intelligence-enabled personalization in interactive marketing: a customer journey perspective. *Journal of Research in Interactive Marketing, (ahead-of-print),* 1-18.

Hu, P., Zeng, Y, Wang, D. & Teng, H. (2024). Too much light blinds: The transparency-resistance paradox in algorithmic management. *Computers in Human Behavior*, 161, https://www.sciencedirect.com/science/article/pii/S0747563224002711.

Isaac, M. S., Wang, R. J., Napper, L. E. & Marsh, J. K. (2024). To err is human: Bias salience can help overcome resistance to medical AI. *Computers in Human Behavior*, 161, https://www.sciencedirect.com/science/article/pii/S074756322400270X.

Kaur, D., Uslu, S., Rittichier, K. J., & Durresi, A. (2022). Trustworthy artificial intelligence: a review, https://dl.acm.org/doi/abs/10.1145/3491209. *ACM Computing Surveys (CSUR), 55*(2), 1-38.

Leschanowsky, A., Rech, S., Popp, B. & Backstrom, T. (2024). Evaluating privacy, security, and trust perceptions in conversational AI: A systematic review. *Computers in Human Behavior*, 161, https://www.sciencedirect.com/science/article/pii/S0747563224002127.

Li, J., Zheng, X., Watanabe, I. & Ochiai, Y. (2024). A systematic review of digital transformation technologies in museum exhibition. *Computers in Human Behavior*, 161, https://www.sciencedirect.com/science/article/pii/S0747563224002759.

Liu, C. & Yin, B. (2024). Affective foundations in AI-human interactions: Insights from evolutionary continuity and interspecies communications. *Computers in Human Behavior*, 161, https://www.sciencedirect.com/science/article/pii/S0747563224002747.

Mayer, C. J., Mahal, J., Geisel, D., Geiger, E. J., Staatz, E., Zappel, M., Lerch, S. P., Ehrenthal, J. C., Walter, S. & Ditzen, B. (2024). *Computers in Human Behavior*, 161, https://www.sciencedirect.com/science/article/pii/S0747563224002875.

Miikkulainen, R. (2024). " Generative AI: An AI paradigm shift in the making?" *AI Magazine* 45: 165–167. https://doi.org/10.1002/aaai.12155

Ognibene, D. et al. (2022). Challenging social media threats using collective well-being-aware recommendation algorithms and an educational virtual companion. Frontiers in Artificial Intelligence 5.

Russell-Rose, T. and Tate, T. (2012). Designing the search experience - the information architecture of discovery. Morgan Kaufmann, ISBN 978-0-12-396981-1, pp. I-XVI, 1-303

Sarkar, A. (2023). Should Computers Be Easy To Use? Questioning the Doctrine of Simplicity in User Interface Design. In *Extended Abstracts of the 2023 CHI Conference on Human Factors in Computing Systems (CHI EA '23), April 23–28, 2023, Hamburg, Germany.* ACM, New York, NY, USA. https://doi.org/10.1145/3544549.3582741

Schmitt, B. H., & Zhang, S. (1998). Language structure and categorization: A study of classifiers in consumer cognition, judgment, and choice. *Journal of Consumer Research, 25*(2), 108-122, https://academic.oup.com/jcr/article-abstract/25/2/108/1799485.

Smale, M. C., Fox, J. D. & Fox, A. K. (2024). When being smart trumps AI: An exploration into consumer preferences for smart vs. AI-powered products. *Computers in Human Behavior*, 161, https://www.sciencedirect.com/science/article/pii/S0747563224002735.

Sundar, A., & Cao, E. S. (2018). Punishing politeness: The role of language in promoting brand trust. *Journal of Business Ethics, 164*, 39-60, https://link.springer.com/article/10.1007/s10551-018-4060-6.

Sundar, A., Cao, E., & Machelit, K. A. (2020). How product aesthetics cues efficacy beliefs of produce performance. *Psychology & Marketing, 37*(9), 1246-62, https://onlinelibrary.wiley.com/doi/abs/10.1002/mar.21355.

Sundar, A., Kardes, F. R., & Wright, S. A. (2015). The influence of repetitive health messages and sensitivity to fluency on the truth effect in advertising. *Journal of Advertising, 44*(4), 375-387, https://doi.org/10.1080/00913367.2015.1045154.

Sundar, A., & Paik, W. (2017). Punishing politeness: Moderating role of belief in just world on severity. *Association for Consumer Research, 45,* 903-905.

Triantafyllopoulos, L., Feretzakis, G., Tzelves, L., Sakagianni, A., Verykios, V. S. & Kalles, D. (2024). Evaluating the interactions of Medical Doctors with chatbots based on large language models: Insights from a nationwide study in the Greek healthcare sector using ChatGPT. *Computers in Human Behavior*, 161, https://www.sciencedirect.com/science/article/pii/S0747563224002723.

Uysal, E., Alavi, S., & Bezençon, V. (2022). Trojan horse or useful helper? A relationship perspective on artificial intelligence assistants with humanlike features. *Journal of the Academy of Marketing Science, 50*(6), 1153-1175.

White, R. W. (2024). Advancing the Search Frontier with AI Agents. Communications of the ACM 67, 9 (September 2024), 54–65. https://doi.org/10.1145/3655615